\documentclass[fleqn,usenatbib]{mnras}
\usepackage{comment}
\usepackage{newtxtext,newtxmath}

\usepackage[T1]{fontenc}

\DeclareRobustCommand{\VAN}[3]{#2}
\let\VANthebibliography\thebibliography
\def\thebibliography{\DeclareRobustCommand{\VAN}[3]{##3}\VANthebibliography}

\usepackage{graphicx}	
\usepackage{amsmath}	
\usepackage{subcaption} 
\usepackage{orcidlink}
\usepackage{xcolor}
\usepackage{soul} 

\newcommand{\deltaF}{$\langle \Delta F \rangle$ }
\newcommand{\rwing}{$R_{\rm wing}$ }
\newcommand\HI{\hbox{H$\,\rm \scriptstyle I$}~}
\newcommand\HII{\hbox{H$\,\rm \scriptstyle II$}~} 

\newcommand\HeII{\hbox{He$\,\rm \scriptstyle II$}~}
\newcommand\HeIII{\hbox{He$\,\rm \scriptstyle III$}~} 
\newcommand\OIII{\hbox{[O$\,\rm \scriptstyle III$]}}
\newcommand\CII{\hbox{[C$\,\rm \scriptstyle II$]}~}
\newcommand{\Lya}{Ly-$\alpha$ } 
\newcommand{\Lyb}{Ly-$\beta$ }

\title[The \Lya damping wing of ULAS J0148+0600]{How probable is the Ly-$\alpha$ damping wing in the spectrum of the redshift \boldmath{z = 5.9896} quasar ULAS J0148+0600?}

\author[F. Sawyer et al.]{Fiona Sawyer$^{1}\orcidlink{0009-0002-2270-5899}$\thanks{E-mail: fiona.sawyer@nottingham.ac.uk},
James S. Bolton$^{1}\,\orcidlink{0000-0003-2764-8248}$,
George D. Becker$^{2}\,\orcidlink{0000-0003-2344-263X}$,
Luke Conaboy$^{1}\,\orcidlink{0000-0002-6580-7177}$,
Martin G. Haehnelt$^{3}\,\orcidlink{0000-0001-8443-2393}$,
\newauthor
Laura Keating$^{4}\,\orcidlink{0000-0001-5211-1958}$,
Girish Kulkarni$^{5}\,\orcidlink{0000-0001-5829-4716}$
and Ewald Puchwein$^{6}\,\orcidlink{0000-0001-8778-7587}$
\\
$^{1}$School of Physics and Astronomy, The University of Nottingham, University Park, Nottingham, NG7 2RD, UK\\
$^{2}$Department of Physics and Astronomy, University of California, Riverside, CA 92521, USA\\
$^{3}$Kavli Institute for Cosmology and Institute of Astronomy, Madingley Road, Cambridge, CB3 0HA, UK\\
$^{4}$Institute for Astronomy, University of Edinburgh, Blackford Hill, Edinburgh, EH9 3HJ, UK\\
$^{5}$Tata Institute of Fundamental Research, Homi Bhabha Road, Mumbai 400005, India\\
$^{6}$Leibniz-Institut f\"ur Astrophysik Potsdam, An der Sternwarte 16, 14482 Potsdam, Germany
}

\date{Accepted XXX. Received YYY; in original form ZZZ}

\pubyear{2025}

\begin{document}
\label{firstpage}
\pagerange{\pageref{firstpage}--\pageref{lastpage}}
\maketitle

\begin{abstract}
The shape of the \Lya transmission in the near zone of the redshift $z=5.9896$ quasar ULAS J0148$+$0600 (hereafter J0148) is consistent with a damping wing arising from an extended neutral hydrogen island in the diffuse intergalactic medium (IGM).   Here we use simulations of late-ending reionisation from Sherwood-Relics to assess the expected incidence of quasars with \Lya and \Lyb absorption similar to the observed J0148 spectrum.  We find a late end to reionisation at $z=5.3$ is a necessary requirement for reproducing a \Lya damping wing consistent with J0148.  This occurs in $\sim3$ per cent of our simulated spectra for an IGM neutral fraction $\langle x_{\rm HI}\rangle=0.14$ at $z=6$.  However, using standard assumptions for the ionising photon output of J0148, the a priori probability of drawing a simulated quasar spectrum with a \Lya damping wing profile \emph{and} \Lya near zone size that simultaneously match J0148 is low, $p<10^{-2}$.  This may indicate that the ionising emission from J0148 is variable on timescales $t<10^{5}\rm\,yr$, or alternatively that the \Lya transmission in the J0148 near zone is impacted by the transverse proximity effect from nearby star-forming galaxies or undetected quasars.  We also predict the IGM temperature should be $T\sim 4\times 10^{4}\rm\,K$ within a few proper Mpc of the \Lya near zone edge due to recent \HI and \HeII photo-heating.  Evidence for enhanced thermal broadening in the \Lya absorption near the damping wing edge would provide further evidence that the final stages of reionisation are occurring at $z<6$.
\end{abstract}

\begin{keywords}
  methods: numerical -- intergalactic medium -- quasars: absorption lines -- dark ages, reionisation, first stars
\end{keywords}


\section{Introduction}
Recent high redshift observations from JWST are challenging our understanding of galaxy formation and the early stages of reionisation \citep[see e.g.,][]{Adamo24}.  The discovery of numerous high redshift galaxies -- now spectroscopically confirmed up to $z \sim 14$ \citep{Adams23, Carniani24} -- that efficiently produce ionising photons has led to a reassessment of the photon budget for reionisation \citep{Atek2024,Simmonds2024,Begley2024,Munoz2024}.   In addition, a large number of intrinsically faint AGN at $z > 5$ \citep{Harikane23, Maiolino23} along with a newly discovered population of 'Little Red Dots' \citep{Labbe23, Matthee24} has led to renewed interest in the role that black hole accretion plays in setting the reionisation photon budget \citep{Dayal24, Madau24, Asthana2024}.

At the same time, much recent progress has been made in our understanding of the final stages of reionisation.  While the first observations of intergalactic \Lya transmission and the \citet{Gunn65} trough in $z\simeq 6$ quasars implied that reionisation was largely complete by $z = 6$ \citep[e.g.,][]{Fan06}, more recent observations -- including notably the XQR-30 survey \citep{d'Odorico23} -- have delivered high quality data that is now firmly challenging this view. This includes confirmation that a fully-ionised intergalactic medium appears to be inconsistent with observed fluctuations in the \Lya forest transmission at $z < 6$  \citep{Becker15, Eilers18, Bosman18, Bosman22}, the rapid evolution of the mean free path of ionising photons at $5 \leq z \leq 6 $ \citep{Becker21, Bosman21, Zhu23, Davies24}, and an incidence of rare dark gaps in the \Lya and \Lyb forest at $z < 6$ that is consistent with at least some fully neutral hydrogen remaining in the diffuse IGM \citep{Becker15, Zhu22, Jin23}.   As speculated by \citet{Lidz2007} and \citet{Mesinger2010}, these data appear to be better reproduced by theoretical models where reionisation ends late ($z \sim 5.3$) and neutral hydrogen islands persist in the diffuse IGM at $z<6$ \citep{Kulkarni19, Keating20a, Keating20b, Nasir20}.

Only very recently, however, has the first possible \emph{direct} evidence for neutral islands in the IGM at $z<6$ been presented.  It was already known that the spectrum of the $z=5.9896$ quasar ULAS J0148$+$0600 (hereafter J0148) contained a remarkably long and dark Ly$\alpha$ trough, with an extent of $\sim 110 h^{-1}$ cMpc \citep{Becker15}.  However, the physical origin of this trough was uncertain; Ly$\alpha$ absorption saturates for neutral hydrogen fractions as small as $x_{\rm HI}\sim 10^{-4}$.  Although some coeval \Lyb transmission spikes were present in the J0148 spectrum presented by \citet{Becker15} -- implying at least some of the long \Lya trough remains highly ionised -- the extended regions with no \Lya and \Lyb transmission could be due to neutral islands or an ionised IGM.  In this context, \cite{Becker24} recently reexamined the Ly$\alpha$ transmission at the red edge of the J0148 trough -- corresponding to the edge of the J0148 \Lya near zone -- using an improved method for removing the quasar continuum.  They determined that the shape of the transmission is consistent with the damping wing profile expected from either an extended neutral island or a compact absorber (i.e., a damped \Lya absorption system, or DLA) at the blue edge of the \Lya near zone.  The DLA interpretation was disfavoured by \citet{Becker24}, however, due to the lack of any corresponding metal lines and/or an associated galaxy.  Subsequently, further weight has been added to the damping wing interpretation by the detection of damping wing-like features in stacked dark gaps at $5.5 \leq z\leq 6$ \citep{Zhu24, Spina24}, following a method originally proposed by \citet{MalloyLidz2015}.

These findings therefore raise the important question of how common the damping wing presented in \citet{Becker24} is within late reionisation models.  While it has only been observed for an individual line of sight in J0148, detailed simulations of inhomogeneous reionisation that resolve the \Lya forest can provide a far larger set of spectra for analysis under a variety of different model assumptions.  In this work we therefore use the Sherwood-Relics simulations of inhomogeneous reionisation \citep{Puchwein23} coupled with a 1D radiative transfer code that follows the ionising emission from quasars \citep{Bolton07} to investigate the likelihood of observing J0148 for a range of different reionisation histories and quasar emission models.   Our approach extends the earlier modelling performed by \citet{Becker24} by using a self-consistent model for patchy reionisation and the local photo-ionisation and heating of the IGM by the quasar.  We also examine the properties of neutral islands in the models that are consistent with the J0148 spectrum, and test the efficacy of the summary statistic used by \citet{Becker24} for identifying the damping wing signature.

This paper is organised as follows. In Section \ref{sec: modeling}, we describe the numerical methods used throughout the paper to create our simulated data set. In Section \ref{sec: probabilities} we discuss the probability that a spectrum with a damping wing signature like J0148 is found within the Sherwood-Relics simulations.  We then examine the physical properties of the IGM around the edge of the quasar near zone in Section \ref{sec:physical}, and assess the \citet{Becker24} summary statistic in Section \ref{sec:efficacy}.  Finally, our conclusions are outlined in Section \ref{sec: conclusions}.


\section{Numerical modelling} \label{sec: modeling}

\subsection{Hydrodynamical simulations of inhomogeneous reionisation}

The cosmological hydrodynamical simulations used in this work are from the Sherwood-Relics suite, described in detail in \cite{Puchwein23}.  The Sherwood-Relics simulations model the IGM in high resolution during and after hydrogen reionisation using a modified version of the cosmological smoothed particle hydrodynamics code P-Gadget-3 \citep{Springel05}.  A novel hybrid radiative transfer (RT) scheme is also used to capture the hydrodynamical effects of inhomogeneous reionisation.   The radiative
transfer is followed using {\sc aton} \citep{Aubert08}, where the
luminosity of the ionising sources are proportional to 
halo mass, and the minimum mass of ionising sources is
$M_{\rm h} > 10^9 h^{-1} \rm{M_{\odot}}$.   The ionising photons have mean energy
$18.6~{\rm eV}$, corresponding to a blackbody spectrum with
temperature $T=4\times10^4~{\rm K}$.  Further technical details may be found in \citet{Puchwein23}.

\begin{table*}
  \caption{Table detailing the properties of each of the simulations used in this work.  From left to right, the table lists: the simulation name used in this work, the box size, the number of particles, the dark matter particle mass, the gas particle mass, the end redshift of reionisation (defined as the redshift where the volume averaged neutral hydrogen fraction first falls below $\langle x_{\rm HI}\rangle=10^{-3}$), the mid-point of reionisation when $\langle x_{\rm HI} \rangle = 0.5$, and finally $ \langle x_{\rm HI} \rangle$ and the volume averaged gas temperature, $\langle T \rangle$, at $z=6.0$. } 
    \centering
    \begin{tabular}{c|c|c|c|c|c|c|c|c}
    \hline
      Name & Box size  & $N_{\rm part}$ & $\rm{M_{dm}}$ & $\rm{M_{gas}}$ & $z_{\rm r}$ & $ z_{\rm mid}$ & $\langle x_{\rm HI}(z=6) \rangle$  & $\langle T(z=6) \rangle$  \\
       & [$h^{-1}$cMpc] & & [$h^{-1}\rm{M_{\odot}}$] &  [$h^{-1}\rm{M_{\odot}}$] & & & & [$10^4$K] \\
       \hline
       160-2048-zr53 & 160 & $2\times2048^{3}$ &3.44 $ \times 10^7$ & 6.38 $\times 10^6$& 5.3 & 7.2 &  0.140 & 10.8 \\
          40-2048-zr53  & 40 & $2\times2048^{3}$ & 5.37 $ \times 10^5$  &  9.97 $\times 10^4$&5.3 & 7.2 &  0.142 & 10.2 \\ 
          40-2048-zr57 & 40 & $2\times2048^{3}$ &5.37 $ \times 10^5$&  9.97 $\times 10^4$&  5.7 & 7.5 & 0.053 & 11.2 \\
     
       \hline
    \end{tabular}
    \label{table:sim_params}
\end{table*}

Here we use the simulations with box sizes 40  $h^{-1}\,\rm cMpc$ and 160 $h^{-1}\,\rm cMpc$, with $2 \times 2048^3$ dark matter and gas particles. The models used are summarised in Table \ref{table:sim_params}.  Our fiducial model uses the 160$h^{-1}\,\rm cMpc$ box as this is better able to capture the large scale structure of the IGM and the sizes of neutral islands / ionised bubbles.  In this model reionisation completes at $z_{\rm r} = 5.3$. This late-ending reionisation history was chosen as it provides good agreement with the distribution of the \Lya forest effective optical depth at $5<z<6$ \citep{Kulkarni19, Bosman22}.   The two 40 $h^{-1}\,\rm cMpc$ models are used to check the assumed simulation mass resolution and to investigate the effect of different reionisation histories on our results, for $z_{\rm r}=5.3$ (matching our fiducial model) and an earlier $z_{\rm r}=5.7$.

\subsection{Radiative transfer simulations of quasar absorption spectra} \label{sec: RT}

The ionising radiation from the quasar is modelled in post-processing using the 1D radiative transfer implementation described in \citet{Bolton07} and most recently used in \citet{Soltinsky23} \citep[see also][for very similar approaches]{Davies2020,Chen2021,Satyavolu2023}.  First, the quasars are placed in the Sherwood-Relics simulations by finding the centre of mass of the 1000 most massive halos at $z = 6$.  We use a friends-of-friends halo finder with linking length $0.2$ times the mean interparticle spacing to identify the haloes.  It is generally expected that at high redshifts luminous quasars are located in large overdensities; \citet{Eilers24} estimate a minimum dark matter host halo mass of $M_{\rm h}\sim 10^{12.43} M_{\odot}$ from the quasar correlation length at $\langle z\rangle=6.25$.   For comparison, in our fiducial simulation (160-2048-zr53) the 1000 most massive dark matter haloes at $z=6$ have masses in the range $10^{11.53}{\rm\,M_{\odot}} \leq M_{\rm h} \leq 10^{12.75} \rm\,M{\odot}$ .  We note, however, that the red edge of the J0148 \Lya trough is $7\rm\,pMpc$ from the quasar.\footnote{Throughout this work, distances are calculated asssuming a quasar systemic redshift of $z=5.9896$, which is based on the \CII emission line redshift (Bosman et al. in prep).}  This is well beyond the dark matter overdensity associated with the host halo and we do not expect the choice of halo mass to significantly impact the \Lya transmission  \citep[see also][]{Keating2015}.

From these haloes, 6000 lines of sight were then extracted (two per spatial dimension, one in each direction) beginning at the centre of mass of the haloes.  Lines of sight were chosen to have length 100 $h^{-1}\,\rm cMpc$, and these sample the gas density, temperature, neutral hydrogen fraction, \HI photoionisation rate and gas peculiar velocity in the hydrodynamical simulations.  Note that to produce a line of sight with the required length from the smaller 40 $h^{-1}\,\rm cMpc$ simulations, two random lines of sight were spliced together with a 20$h^{-1}\,\rm cMpc$ skewer starting at the centre of mass of a halo (accounting for appropriate redshift evolution).  

The 1D radiative transfer code was then used to simulate the presence of a quasar with absolute magnitude $M_{1450}$ and lifetime $t_{\rm q}$. A double power law spectrum was used for the quasar spectral energy distribution, of the form 
\begin{equation}
    L_{\nu} = \left\{
        \begin{array}{ll}
            \nu^{-0.61} & \quad ( 1050\, \text{\AA}\, < \lambda < 1450\, \text{\AA}), \\
            \nu^{-\alpha_{\rm s}} & \quad (\lambda < 1050\, \text{\AA}),
        \end{array}
    \right.
    \label{eq:quasar luminosity}
\end{equation}
where $\alpha_{\rm s} = 1.5$ for our fiducial model \citep{Lusso15,Shen20}. This spectrum was then normalised to an absolute magnitude at $1450 \text{\AA}$, $M_{1450} = -27.4$, which is the absolute magnitude for J0148 reported by \cite{Eilers17}.  This results in an ionising photon emission rate of $\dot{N} = 3.0 \times 10^{57} \rm{s^{-1}}$.  Our fiducial model assumption for the optically/UV bright lifetime of the quasar is $t_{\rm q}=10^{7}\rm\,yr$.

\begin{figure*}
    \includegraphics[trim = 0cm 2cm 2cm 4cm, clip, scale = 0.5]{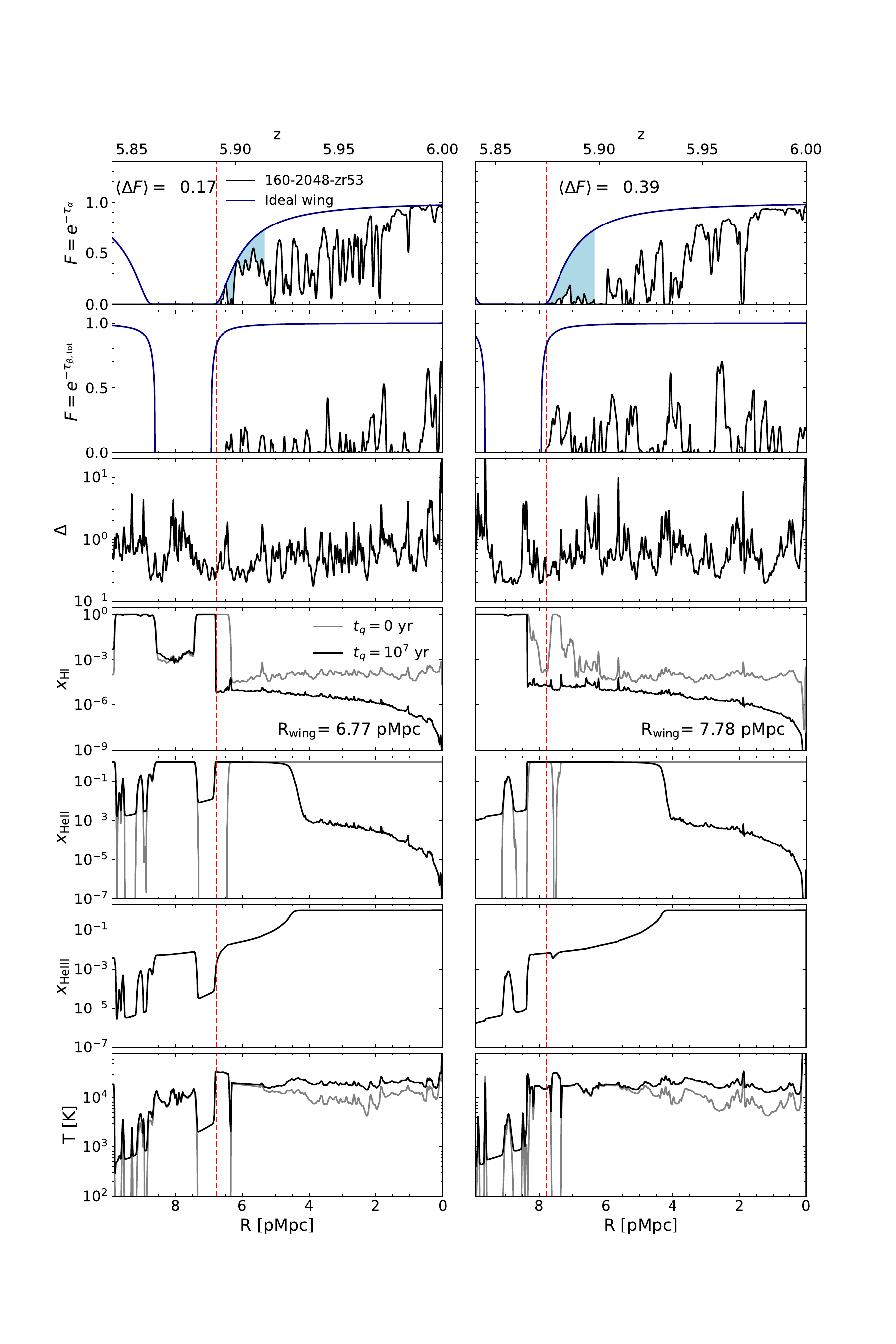}
    \vspace{-1.0cm}
\caption{{\it Top panels:} Example radiative transfer simulations of Ly-$\alpha$ transmission in the near zone of a bright quasar at $z=6$, drawn from our fiducial quasar model with $M_{1450}=-27.4$, $t_{\rm q}=10^{7}\rm\,yr$, $\alpha_{\rm s}=1.5$ and an initial volume weighted IGM neutral hydrogen fraction $\langle x_{\rm HI}\rangle = 0.14$.  The left panel shows a sight-line with $\langle \Delta F \rangle=0.17$ (shaded blue region), similar to the value of $\langle \Delta F \rangle =0.15$ for J0148$+$0600 reported by \citet{Becker24}. The right panel displays a more typical sight-line from our simulated data set with $\langle \Delta F\rangle =0.39$.  The blue solid curve shows the damping wing template from a $7.5h^{-1}\rm\,cMpc$ neutral island, and the red dashed line gives $R_{\rm wing}$, defined here as the distance in proper Mpc from the quasar where the damping wing template first exceeds $F=0.01$. 
{\it Second panels:} The Ly-$\beta$ transmission with the ideal damping wing.
{\it Third panels:} Overdensity $\Delta$ along the simulated line of sight.
{\it Fourth panels:} The corresponding neutral hydrogen fraction, $x_{\rm HI}=n_{\rm HI}/n_{\rm H}$, along the simulated sight-lines.  The grey curves show $x_{\rm HI}$ prior to the quasar turning on.  Note the presence of a neutral island just beyond the \HII ionisation front in the left panel at $R=6.8\rm\,pMpc$.
{\it Fifth panels:} The \HeII fraction, $x_{\rm HeII}=n_{\rm HeII}/n_{\rm He}$, with line colours matching the panel above.
{\it Sixth panels:} The \HeIII fraction, $x_{\rm HeIII}=n_{\rm HeIII}/n_{\rm He}$, along the line of sight. Note that prior to the quasar turning on, $x_{\rm HeIII}$ is assumed to be zero. 
{\it Bottom panels:} Temperature of the IGM at the time of observation and prior to the quasar turning on. A region of enhanced temperature associated with the photo-heating of the recently reionised region near \rwing is evident in the left panel.
}
    \label{fig:example_los}

\end{figure*}

After including the photo-ionisation\footnote{More specifically, the photo-ionisation rate from the quasar is \emph{added} to the initial photo-ionisation rates used in the Sherwood-Relics patchy reionisation model.} and photo-heating of the IGM by the quasar using the 1D radiative transfer, the \Lya optical depths along the line of sight were calculated assuming a Voigt line profile \citep{Tepper-Garcia06}.  The corresponding \Lyb forest optical depths were also calculated from the sum of the \Lyb optical depths (derived by scaling the \Lya optical depths) and a foreground \Lya contribution beginning at a redshift $z^{\prime}$, given by
\begin{equation}
    z^{\prime}=\frac{\lambda_{\beta}(1+z)}{\lambda_{\alpha}}-1,
\end{equation}
where $z$ is the redshift of the quasar and $\lambda_{\beta}$ and $\lambda_{\alpha}$ are the \Lya and \Lyb rest frame wavelengths. The total \Lyb optical depth, $\tau_{\rm{Ly\beta, tot}}$, is then given by
\begin{equation}
\centering
    \tau_{\beta, \rm tot}(z) = \tau_{\alpha, \rm{fore}}(z^{\prime}) + \tau_{\alpha}(z) \frac{f_{\beta} \lambda_{\beta}}{f_{\alpha} \lambda_{\alpha}},
\end{equation}
where $f_{\beta}$ and $f_{\alpha}$ are the \Lya and \Lyb oscillator strengths and $\tau_{\alpha, \rm{fore}}$ is the \Lya forest optical depth at $z^{\prime}$.  In this case, the foreground \Lya spectrum is created from random lines of sight drawn from the Sherwood-Relics simulation at $z^{\prime}$.  In order to make the simulated spectra comparable to the spectrum of J0148, the transmitted flux was then rebinned with pixel size of 10 $\rm km\,s^{-1}$ and convolved with a Gaussian profile with FWHM = 23 $\rm km\,s^{-1}$, appropriate for the X-Shooter spectrograph on the Very Large Telescope \citep{d'Odorico23}.

Finally, to investigate how the quasar properties may impact on the probability of finding a damping wing similar to that observed in J0148, additional quasar models were created where the quasar luminosity, extreme UV spectral index $\alpha_{\rm s}$ and optically bright lifetime were changed from the fiducial model described above. To observe the impact of the simulation box size and reionisation history, each of the 40$h^{-1}\rm\,cMpc$ box simulations described in Table \ref{table:sim_params} were also examined.  In this work, $\alpha_{\rm s}$ is varied as 0.5, 1.5 or 2.5, the absolute magnitude is varied between $M_{1450}=-26.65$ and $M_{1450}=-28.9$ (0.5 to 4 $L_q$), and quasar lifetimes between $10^5\rm\, yr$ \citep[cf.][]{Morey21} and $10^8$ yr are considered.  Note that we only consider a ``light bulb'' quasar model in this work, and do not model more complicated quasar light curves \citep[e.g.][]{Davies2020,Satyavolu2023,Soltinsky23, Zhou24}.  We shall discuss the possible implications of this choice later.

\subsection{The damping wing in simulated quasar spectra} \label{sec:idealwing}

\begin{figure}
    \centering
\vspace{-0.4cm}
    \includegraphics[scale = 0.6]{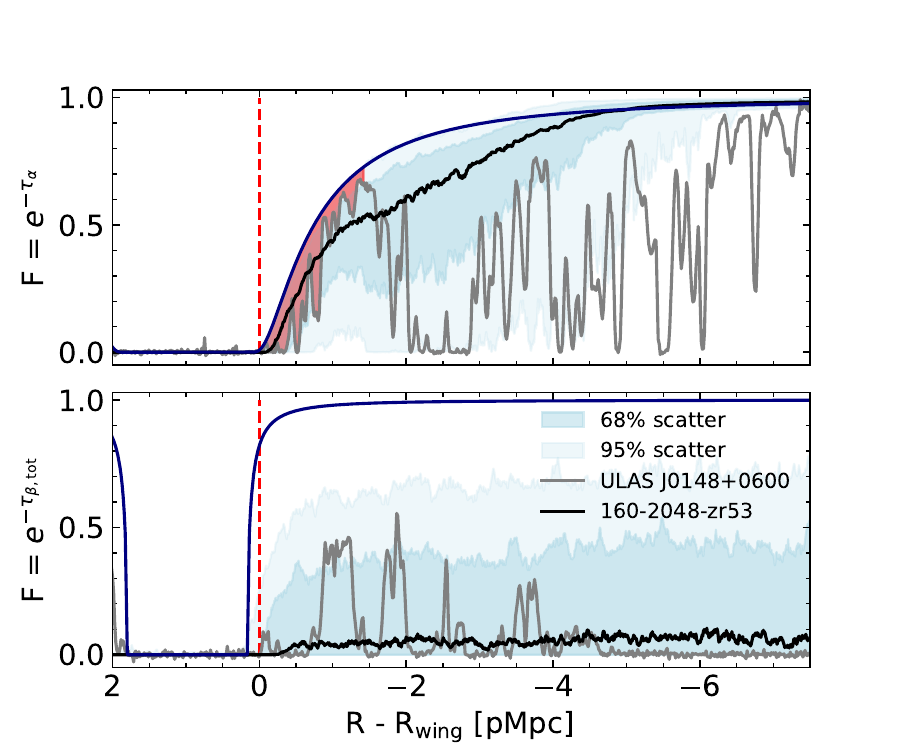}
    \vspace{-0.6cm}
    \caption{The stacked \Lya (top panel) and \Lyb (bottom panel) transmission about $R-R_{\rm wing}=0$ (red dashed line) for our fiducial model. The black curves show the median transmission for lines of sight with \deltaF $= 0.15\pm 0.05$ while the light blue shaded regions show the $68$ and $95$ per cent scatter. The observed transmission from J0148$+$0600 is shown in grey, with a damping wing template (see text for details) with \rwing = 0 pMpc shown by the navy curve. The red dashed line at \rwing is also shown as a reference, and the shaded red region shows the difference between the observed spectrum and the ideal damping wing over the 1000 $\rm{kms^{-1}}$ region used to calculate \deltaF.}
    \label{fig: wing transmission}
\end{figure}

Figure \ref{fig:example_los} presents two example quasar spectra produced from our fiducial model following the methods described above. For each spectrum we show the \Lya and \Lyb transmission (top and second panels) as well as the gas overdensity (third panels), neutral hydrogen fraction $x_{\rm{HI}}$ (fourth panels), singly ionised helium fraction $x_{\rm{HeII}}$ (fifth panels), doubly ionised helium fraction $x_{\rm{HeIII}}$ (sixth panels) and gas temperature (bottom panels) along the line of sight for an optically/UV bright lifetime of $t_{\rm q}= 10^7$yr.  The grey curves show the quantities before the quasar turns on. 

Each example spectrum in Figure \ref{fig:example_los} also has an idealised damping wing template (blue curves) and the value of the summary statistic \deltaF introduced by \citet{Becker24} is reported.  In brief, \citet{Becker24} obtained \deltaF by creating a damping wing template and calculating the mean difference between the \Lya transmission and the template (shown as the shaded region in Figure~\ref{fig:example_los}).  For J0148, \citet{Becker24} reported a value of \deltaF $ = 0.15$, with smaller values of this statistic indicating closer agreement between the template and the \Lya transmission.  In \citet{Becker24} and in this work, the template is created assuming a region of neutral hydrogen at the mean background density with size 7.5 $h^{-1}\,\rm cMpc$.  The optical depths for the template are calculated using the Voigt profile approximation from \citet{Tepper-Garcia06}. The extent of the neutral region is set by the size of the dark region in the \Lya and \Lyb forest of J0148 at the blue edge of the \Lya near zone (see Figure 2 in \cite{Becker24}), although this choice did not have a significant effect on the results reported by \citet{Becker24}.    Finally, the placement of the damping wing template is set such that the difference between the template and the \Lya transmission is minimised over a range of $1000\rm\,km\,s^{-1}$, starting from the point where the \Lya damping wing template first exceeds $F=0.01$.   The \Lya transmission in the quasar spectrum must also not exceed the ideal damping wing template over this range.  The choice of the range we use to calculate \deltaF does not significantly impact our results, as also found in \citet{Becker24}. Appendix \ref{app:region size} provides a more detailed discussion of the impact of changing the range over which we calculate \deltaF.

In this work we add one further criterion when analysing our models.  To ensure that the damping wing template has a dark region blueward of the wing edge, the dark trough with extent $7.5h^{-1}\,\rm cMpc$ must also have transmission $F \leq 2/{\rm{SNR}}$ in \emph{both} the simulated \Lya and \Lyb spectra. The maximum allowed transmission in the dark region was chosen from the average signal to noise ratio (SNR) of the J0148 X-Shooter spectrum, SNR$=59.9$ \citep{d'Odorico23}. If a dark trough of the required length does not exist in the simulated spectrum, then the damping wing template fit fails and a value of \deltaF is not returned.  In addition, we use the damping wing template as an approximate measure of the quasar \Lya near zone size, with the distance between the quasar and where the \Lya template first exceeds $F=0.01$ referred to hereafter as $R_{\rm wing}$.   The edge of the near zone using this definition is shown in Figure \ref{fig:example_los} as a red dashed line, with \rwing also reported. A flow chart offering a more detailed overview of the algorithm used to calculate \deltaF and $R_{\rm wing}$ is provided in Appendix \ref{app:flowchart}.

The left panels of Figure \ref{fig:example_los} show a spectrum similar to that of J0148, which has \deltaF = 0.15 and \rwing = 7.00 pMpc \citep{Becker24}. A neutral island is still present at the edge of the quasar \Lya near zone for $t_{\rm q}=10^{7}\rm\,yr$.  In contrast, the right panels instead show a spectrum with \deltaF and \rwing values that are more typical for our fiducial model.  The small neutral island at $R\sim 7.5\rm\,pMpc$ present before the quasar turns on has been fully ionised by the quasar radiation field.\footnote{For a quasar age of $t_{\rm q}=10^{7}\rm\,yr$, the distance travelled by a photon is $R=ct_{\rm q}\sim 3\rm\,pMpc$, yet the quasar \HII ionisation front in Fig.~\ref{fig:example_los} is instead at $R\sim 7\rm\,pMpc$.  This is because throughout this work we assume the quasar photons are observed at the same retarded time, $t_{\rm R}=t-R/c$, where $t$ is the time when the photons reach a distance $R$ from the quasar.  From the observers' perspective, therefore, the \HII ionisation front (and \Lya near-zone) will appear to undergo a superluminal expansion phase when the number of ionising photons per hydrogen atom significantly exceeds unity \citep[see e.g. Appendix A in][]{Bolton07}.} There is also a large neutral island shortly after \rwing for this line of sight, but its presence does not result in a small \deltaF value because of the low level of transmission at the near-zone edge. As we discuss later in Sections \ref{sec:physical} and \ref{sec:efficacy}, the presence of a neutral island close to \rwing does not guarantee a low \deltaF value.  Reproducing the J0148 spectrum instead requires a rare combination of physical conditions to hold for the IGM at the blue edge of the near-zone.  

For a more direct comparison to the J0148 spectrum, in Figure \ref{fig: wing transmission} we show the median \Lya and \Lyb transmission with the 68 and 95 per cent scatter for all spectra in our fiducial model with \deltaF $= 0.15 \pm 0.05$.  Note the simulated spectra are all stacked around $R - R_{\rm wing}=0$ (red dashed line). This causes the scatter to be suppressed in the \Lya stack as $R - R_{\rm wing}$ becomes more negative; the near-zone size in each of the simulated lines of sight in the stack is different, and most lines of sight have a value of \rwing smaller than 7 pMpc.  At the near-zone edge, the simulated spectra are nevertheless broadly consistent with the observed J0148 transmission within the scatter, shown as the grey curve in Figure \ref{fig: wing transmission}.


\section{The probability of the J0148+0600 damping wing in Sherwood-Relics}
\label{sec: probabilities}
\subsection{The distribution of \deltaF}

\begin{figure*}
\centering
    \includegraphics[scale = 0.8]{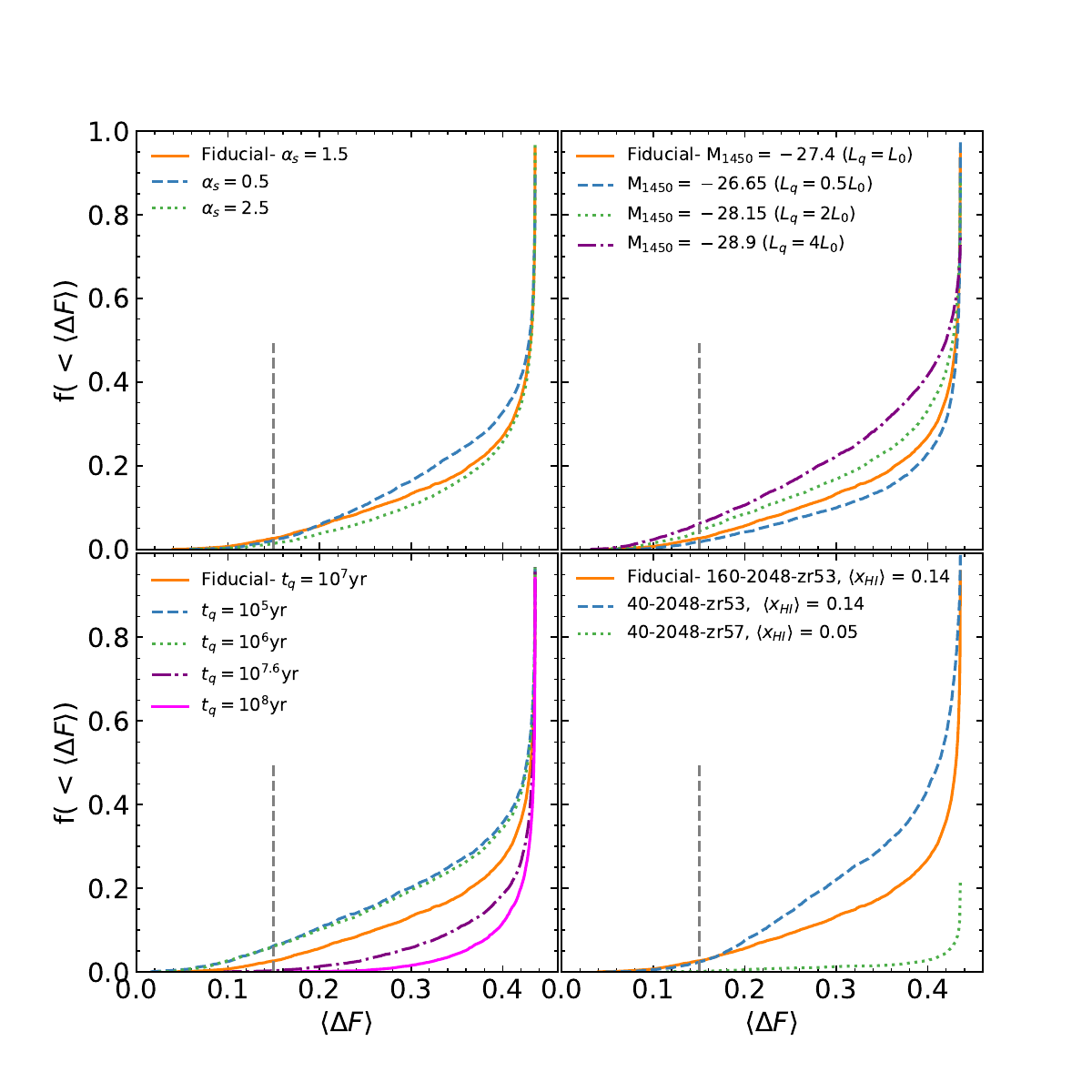}
  \vspace{-1.0cm}
    \caption{The distributions of \deltaF predicted by our radiative transfer simulations of Ly-$\alpha$ transmission from the IGM around quasars at $z=6.0$.  Each distribution is constructed from $6000$ unique sight-lines.   The grey dashed line shows the observed value of \deltaF $=0.15$ for ULAS J0148$+$0600 \citep{Becker24}.  Each panel shows the impact of varying a different parameter in the simulations.  Clockwise from the top left, these are: 
    the extreme UV spectral index, $\alpha_s$, of the quasar; the absolute UV magnitude of the quasar, $M_{1450}$, selected here to match the luminosity scalings investigated by \citet{Becker24}; simulation box size/mass resolution and the volume averaged IGM neutral hydrogen fraction prior to the quasar turning on, and the optically/UV bright lifetime of the quasar, $t_{\rm q}$, assuming a "light-bulb" emission model.  In general, we expect more sight-lines with $\Delta F\leq 0.15$ for either brighter UV absolute magnitudes/luminosities, shorter optically/UV bright lifetimes, and larger initial IGM neutral hydrogen fractions.  In contrast, the fraction of sight-lines  with $\Delta F\leq 0.15$ is not particularly sensitive to either the EUV spectral index or simulation box size/mass resolution.}
    \label{fig:CDFs}
\end{figure*}

\begin{table*}
      \caption{Probabilities of drawing a J0148-like line of sight from our models.  Each model variation consists of 6000 individual 1D radiative transfer simulations.  Columns 1-5 summarise the quasar model parameters, which are from left to right: the base hydrodynamical simulation used, the initial volume averaged \HI fraction, the quasar absolute UV magnitude, the extreme-UV spectral index of the quasar, and the optically/UV bright lifetime of the quasar.  The fiducial model is in bold. Columns 6 and 7 list the a priori probabilities for drawing a J0148-like spectrum, for either $\langle \Delta F \rangle \leq 0.15$ or for a joint constraint of $\langle \Delta F \rangle = 0.15 \pm 0.05$ and $ R_{\rm{wing}} = 7.00 \pm 0.5\rm\,pMpc$. }
    \begin{tabular}{c|c|c|c|c|c|c}
    \hline
      Hydrodynamical  & $\langle x_{\rm HI} \rangle$ & $M_{1450}$  & $\alpha_{\rm s}$ & $t_{\rm q}$ [yr]  &  $p(\langle \Delta F \rangle \leq 0.15)$  &  $p(\langle \Delta F \rangle = 0.15 \pm 0.05\, \land $ \\
 
      simulation name & &  & &   &  &   $R_{\rm{wing}} = 7.00 \pm 0.5\rm\,pMpc)$ \\
    \hline 
    {\bf 160-2048-zr53} & {\bf 0.140} & {\bf -27.4} & {\bf 1.5} & \boldmath{$10^7$} & {\bf 0.0283} & {\bf0.0002} \\
    160-2048-zr53 & 0.140 & -27.4 & 0.5 & $10^7$ & 0.0250& 0.0247 \\
    160-2048-zr53 & 0.140 & -27.4 & 2.5 & $10^7$ & 0.0152 & 0.0\\
    160-2048-zr53 &  0.140 & -27.4 & 1.5 & $10^5$ & 0.0665& 0.0 \\
    160-2048-zr53 &  0.140 & -27.4 & 1.5 & $10^6$& 0.0642 & 0.0 \\
    160-2048-zr53 &  0.140 & -27.4 & 1.5 & $10^{7.6}$ & 0.0033& 0.0080\\
    160-2048-zr53 &  0.140 & -27.4 & 1.5 & $10^8$ & 0.0 & 0.0\\
    160-2048-zr53 & 0.140 & -26.65 & 1.5 & $10^7$& 0.0195& 0.0\\
    160-2048-zr53 & 0.140 & -28.15 & 1.5 & $10^7$& 0.0475 & 0.0135\\  
    160-2048-zr53 & 0.140 & -28.9 & 1.5 & $10^7$ & 0.0648 & 0.0260\\
    40-2048-zr53 & 0.142 & -27.4 & 1.5 & $10^7$& 0.0270& 0.0 \\
    40-2048-zr57 & 0.053 & -27.4 & 1.5 & $10^7$& 0.0045& 0.0006\\
    \hline
    
    \end{tabular}

    \label{tab:model_comparison}
\end{table*}

\begin{figure*}
\centering
    \includegraphics[scale = 0.65]{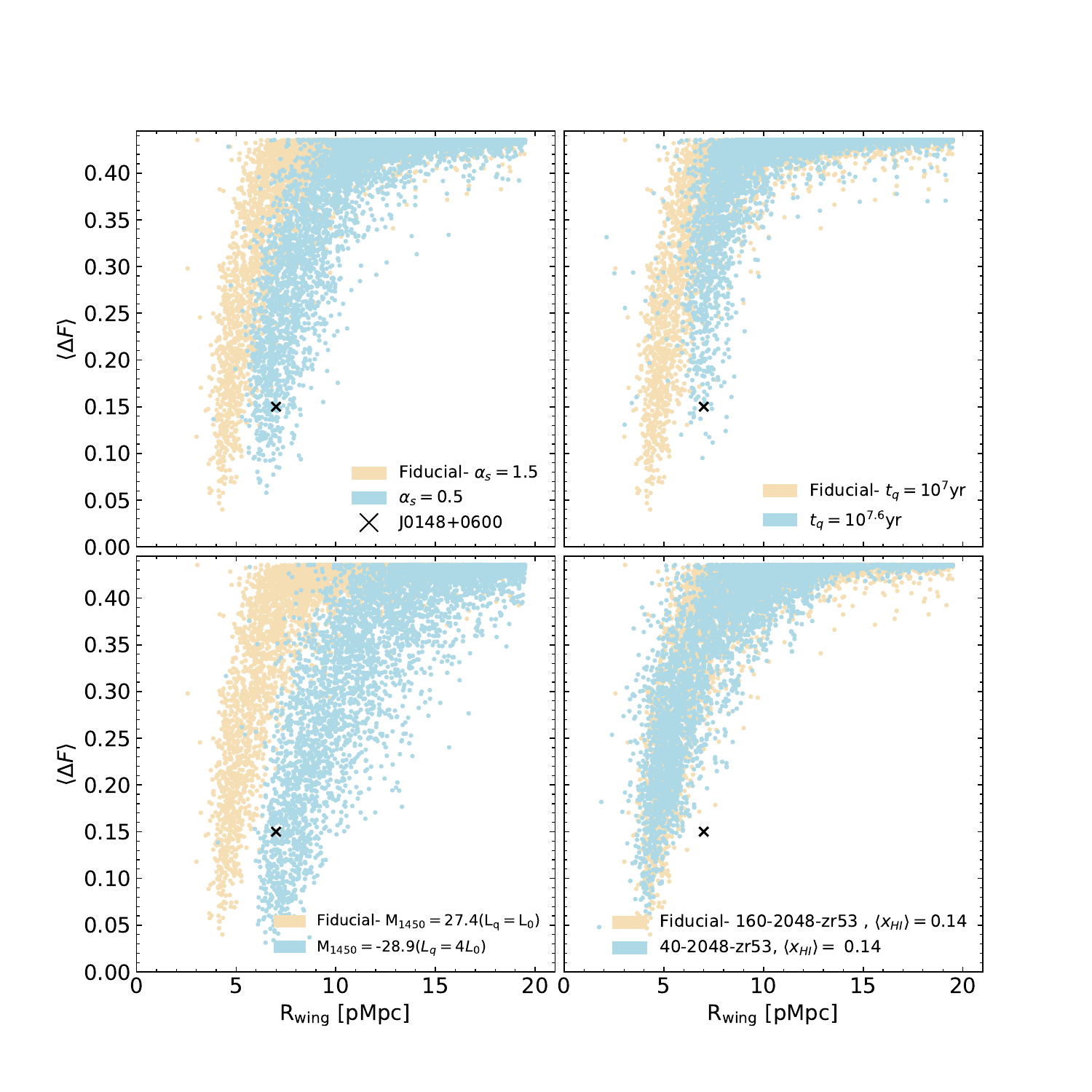}
    \vspace{-1.0cm}
          \caption{Scatter plots of \deltaF against $R_{\rm wing}$.  All sight-lines that return a value of \deltaF are displayed. Clockwise from top left, the fiducial model (orange) is compared to: a model with a harder extreme UV spectral index, $\alpha_{\rm s}=0.5$ (blue); a model with a longer optically/UV bright quasar lifetime, $t_{\rm q}=10^{7.6}\rm\,yr$; a model with a brighter UV absolute magnitude, $M_{1450}=-28.9$; a model constructed from a hydrodynamical simulation with box size $40h^{-1}\rm\,cMpc$ and $64$ times better resolution, for the same initial volume averaged IGM neutral fraction.  The parameters for J0148+0600 are given by the black cross \citep{Becker24}.   The a priori probability of a J0148-like spectrum with $\langle \Delta F \rangle = 0.15 \pm 0.05$ and $ R_{\rm{wing}} = 7.00 \pm 0.5\rm\,pMpc$ in our fiducial model is $p<10^{-3}$.  In general, simulations with a harder EUV spectral index, longer optically/UV bright lifetime, or brighter UV absolute magntitude are better able to reproduce both the observed \deltaF \emph{and} $R_{\rm wing}$, although these sight-lines remain rare, $p<0.03$ (see Table~\ref{tab:model_comparison}). }
    \label{fig: scatterplot}
\end{figure*}

We now turn to examining the likelihood of obtaining a J0148-like spectrum from our simulations, by performing the damping wing template fitting procedure and finding $\langle \Delta F \rangle $ for the 6000 lines of sight in each of the models. We first calculate the fraction of lines of sight with \deltaF below some value, $f(<\langle \Delta F \rangle)$. Figure \ref{fig:CDFs} shows the results for each model.  Note that in Figure \ref{fig:CDFs} the distributions do not all reach $f(<\langle \Delta F \rangle)=1$.  This is because some of the 6000 lines of sight do not meet the criteria for fitting the damping wing template described in Section~\ref{sec:idealwing} and therefore do not return a value for $\langle \Delta F \rangle$.  This is most noticeable for the 40-2048-zr57 model in the lower right panel of Figure~\ref{fig:CDFs}, where almost 80 per cent of the parent sample fails to return a $\langle \Delta F \rangle$ value.

Each panel in Figure \ref{fig:CDFs} shows the effect of changing one model parameter from the fiducial model, clockwise from top left: the extreme UV (EUV) spectral index, $\alpha_{\rm s}$, of the quasar; the absolute UV magnitude of the quasar, $\rm{M_{1450}}$, selected here to match the luminosity scalings investigated by \cite{Becker24}; the simulation box size / mass resolution and the (initial) volume averaged IGM neutral hydrogen fraction\footnote{We also explored a $40h^{-1}\rm\,cMpc$ simulation with reionisation ending at $z_{\rm r}=6$ with $\langle x_{\rm HI}(z=6) \rangle=0.002$.  However, in this model with a highly ionised IGM at $z=6$ there were no lines of sight that met our criteria for fitting the damping wing template. We therefore do not consider this model further here.} and the the optically/UV bright lifetime of the quasar, $t_{\rm q}$. On each panel the fiducial model is shown for comparison as an orange curve.  The vertical dashed line displays the measured $\langle \Delta F \rangle =0.15$ for J0148.

In Table ~\ref{tab:model_comparison} we show the probabilities that quasar spectra in each model have \deltaF $\leq 0.15$, drawn from the parent population of 6000 sight lines for each model. Quasar spectra with \deltaF $\leq 0.15$ are rare; for the fiducial model $p(\langle \Delta F \rangle \leq 0.15) = 0.028$, and for all of the models $p(\langle \Delta F \rangle \leq 0.15) <0.100$. The quasar luminosity, lifetime and the neutral hydrogen fraction of the simulation all have an impact on $p(\langle \Delta F \rangle \leq 0.15)$; brighter quasars, shorter lifetimes and higher neutral fractions all increase the likelihood of observing a spectrum with a \deltaF similar to J0148.   In all these cases this is because the \Lya transmission close to the edge of the quasar near zone is maximised and so the damping wing template returns smaller \deltaF values.  A brighter quasar will have an increased emission rate of ionising photons, $\dot{N}$, such that the \HI photo-ionisation rate, $\Gamma_{\rm HI}$, at the edge of the quasar near zone will be larger.  For the shorter quasar lifetimes the size of the near zone is smaller, and this also results in a larger \HI photo-ionisation rate at the near zone edge (in the optically thin \HII region, $\Gamma_{\rm HI}\propto R^{-2}$). Lastly, an IGM with a higher neutral fraction also means the near zones are typically smaller.  The effect of the volume averaged neutral hydrogen fraction on the probability of observing low \deltaF values is also pronounced, with the J0148 \deltaF very unlikely for reionisation ending at $z_{\rm r} = 5.7$ (see Table~\ref{tab:model_comparison} for the tabulated probabilities).  This demonstrates that the damping wing signature and its incidence rate in the $z=6$ quasar population is sensitive to the IGM neutral fraction, as many earlier studies have demonstrated  \citep[e.g.][]{Mesinger2007,Mortlock2011,Bolton2011, Davies2018,Wang2020,Greig2022,Durovcikova2024,Hennawi2024}

In contrast, the EUV spectral index has a limited effect on $p(\langle \Delta F \rangle \leq 0.15)$. Likewise, the combined effect of the simulation box size and mass resolution, as demonstrated by comparing the two simulations with the same neutral fractions but differing box sizes in the lower right panel of Figure~\ref{fig:CDFs}, only begins to affect \deltaF $\geq 0.2$.  We explore the effect of mass resolution on the distribution of \deltaF further in Appendix \ref{app:converge}.  In summary, we expect the mass resolution of our fiducial 160-2048-zr53 simulation should not significantly impact our assessment of $p(\langle \Delta F \rangle \leq 0.15)$ \citep[see also appendix B in][]{Becker24}, and that lowering the mass resolution of the hydrodynamical simulation only makes J0148-like spectra \emph{more likely} in our models.

\subsection{The joint distribution for \deltaF and $R_{\rm wing}$}

Although we have demonstrated that the probability of obtaining a spectrum with $\langle \Delta F \rangle \leq 0.15$ -- matching J0148 -- is small, this comparison ignores the extent of the J0148 near zone as measured by $R_{\rm wing}$. Hence, in addition to investigating probability of finding a quasar with the value of \deltaF observed from the spectrum of J0148, we now consider the probability of obtaining both the \deltaF value and \Lya near zone size, $R_{\rm wing}$, observed for J0148.   

The numerical results are again summarised in Table \ref{tab:model_comparison}, while Figure~\ref{fig: scatterplot} displays the joint distribution for \deltaF and \rwing for the fiducial model (orange points) and selected alternative models (blue points) where we again vary the EUV spectral index of the quasar, the quasar lifetime, the absolute UV magnitude of the quasar and the simulation box size / mass resolution.  The probability that a line of sight has \emph{both} $\langle \Delta F \rangle = 0.15 \pm 0.05$ and $R_{\rm wing}=7.0 \pm 0.5 \rm\,pMpc$ (shown as the black cross in Figure~\ref{fig: scatterplot}) is always smaller than requiring $p(\langle \Delta F \rangle \leq 0.15)$.   Note also that spectra with lower \deltaF values tend to have smaller \rwing values, but for $\langle \Delta F \rangle >0.4$ the value of \rwing becomes independent of $\langle \Delta F \rangle$. Interestingly, the quasar models from which we are most likely to observe a J0148-like spectrum now occur for a higher luminosity / absolute magnitude $M_{1450}$ (4$L_0$), or a harder quasar spectrum ($\alpha_s = 0.5$), or for a longer optically/UV bright lifetime ($t_{\rm q} = 10^{7.6}\rm\,yr$).   The higher luminosity or harder spectrum both increase the $\dot{N}$ of the quasar compared to the fiducial model, and so are capable of generating larger \rwing values.  Similarly, a longer optically/UV bright lifetime can extend the \Lya near zone around the quasar.

However, it seems unlikely that the EUV spectral slope of J0148 will be significantly harder than the composite quasar continuum used in \citet{Becker24}, or that the UV absolute magnitude is significantly in error. 
\citet{Morey21} find a typical optically/UV bright lifetime of $t_{\rm q} \sim 10^{6}\rm\,yr$ in their analysis of $z\sim 6$ quasar near zone sizes, which is over an order of magnitude shorter than the model that reproduces the J0148 spectrum most readily.  On the other hand, for 22 quasars from the XQR-30 survey \citet{Satyavolu2023_XQR30} reported a broad range of near-zone sizes at $z\sim 6$ that are consistent with optically/UV bright lifetimes of $t_{\rm q}\sim 10^{4}$--$10^{8}\rm\,yr$.  A quasar lifetime of $t_{\rm q} = 10^{7.6}$ yr could therefore partially explain the J0148 spectrum, although the probability of producing a damping wing similar to J0148 is still small, $p < 10^{-2}$.

This suggests that either (i) quasars with \Lya near zone absorption profiles and giant \Lya absorption troughs similar to J0148 at $z=6$ are rare (with a priori probabilities $p<10^{-2}$), (ii) the damping wing in J0148 is instead due to an extremely low metallicity DLA\footnote{Note that we do not model compact absorbers with damping wings in our simulations, and are therefore unable to directly assess how likely this is.}, although as mentioned earlier this was already argued against by \citet{Becker24} or (iii)  some additional physical effect is missing from our models.   In the latter case, one possibility is that J0148 may have a variable light curve \citep[see e.g.][]{Davies2020,Satyavolu2023,Soltinsky23} and until recently it was much brighter.  This would require variability on a timescale comparable to or shorter than the equilibriation timescale, $t_{\rm eq}$, for the \HI in the \Lya near zone, where
\begin{equation} t_{\rm eq} = \frac{x_{\rm HI,\rm\,eq}}{n_{\rm e} \alpha_{A}(T)} \simeq \frac{10^{5.0}\rm\, yr}{\Delta} \left(\frac{x_{\rm HI,\rm\,eq}}{10^{-4}}\right)\left(\frac{T}{10^{4}\rm\,K}\right)^{0.72}\left(\frac{1+z}{7}\right)^{-3}. \label{eq:teq} \end{equation}
Here $x_{\rm HI,\,eq}$ is the \HI fraction in ionisation equilibrium, $\alpha_{\rm A}=4.06\times 10^{-13}\mathrm{\,cm^{3}\,s^{-1}}(T/10^{4}\rm\,K)^{-0.72}$ is the case-A recombination coefficient and $n_{\rm e}=1.158 n_{\rm H}$ for fully ionised hydrogen and helium.   Alternatively, as already noted by \citet{Becker24}, another possibility is the transverse proximity effect.  Enhanced ionisation may occur if there are additional ionising sources near the location of the blue edge of the J0148 \Lya near zone \citep[see e.g.][for possible candidates at $z\simeq 5.92$]{Eilers24}. The sources could also potentially be undetected quasars, which are expected to cluster on the relevant scales \citep{Shen07, Eilers24} but may not be detectable due to the anisotropic nature of quasar emission, particularly if the quasar is intrinsically faint.  Quasar variability and/or duty cycle could also play a role in limiting detectability.


\section{Physical properties of gas associated with the damping wing} \label{sec:physical}

\begin{figure*}
\centering
\includegraphics[trim=1.7cm 0cm 0cm 0cm, clip, scale = 0.68]{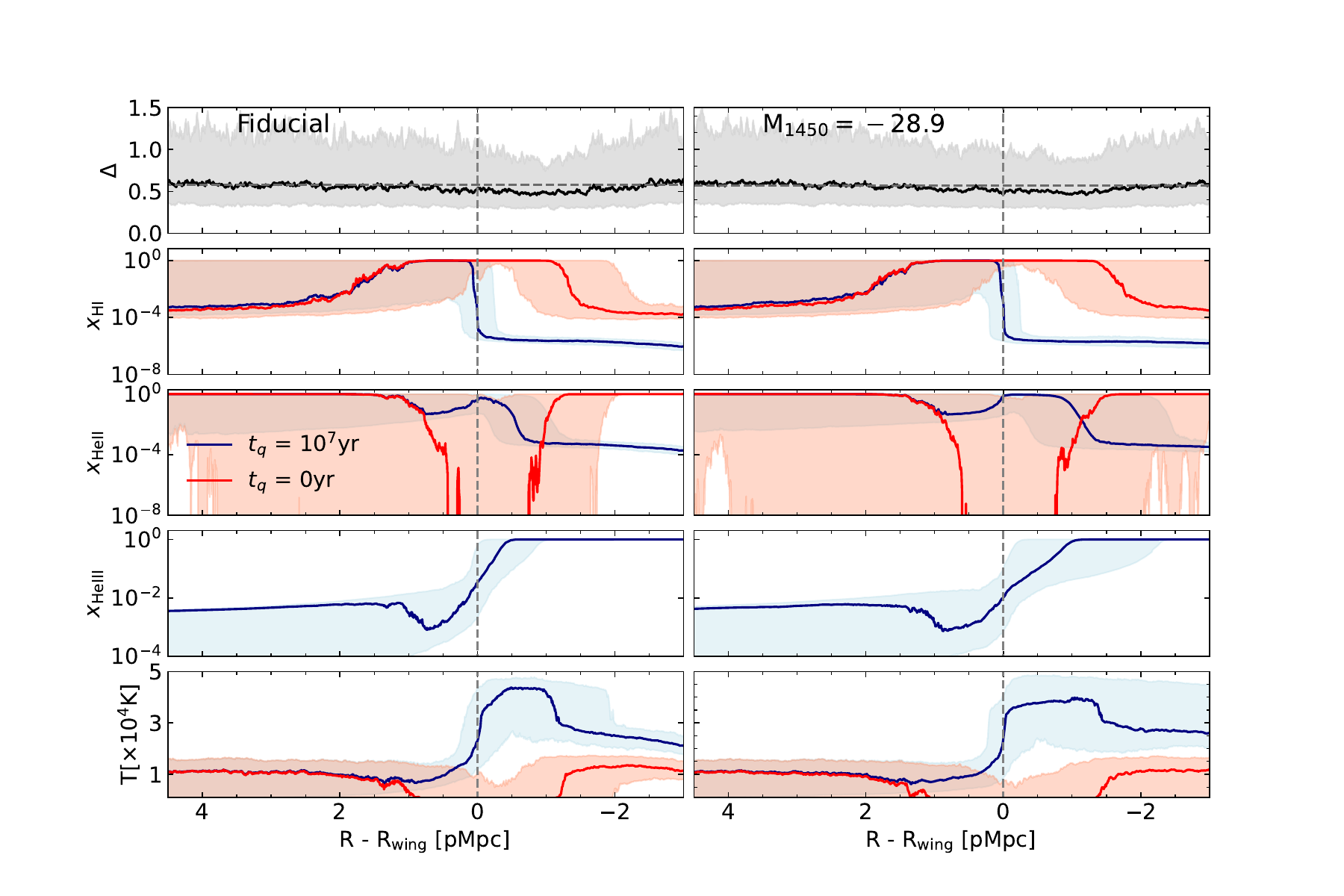}
\vspace{-0.8cm}
\caption{The median line of sight gas density, $\Delta=\rho/\langle \rho \rangle$, \HI fraction, \HeII fraction, \HeIII fraction and gas temperature, for the sub-set of sight-lines  with $\langle \Delta F \rangle = 0.15 \pm 0.05$ drawn from a total of $6000$ simulated spectra.  The left column shows the fiducial model (averaged over 296 sight-lines), while the right column shows the $M_{1450}=-28.9$ model (averaged over 493 sight-lines).  The distances on the horizontal axes have been renormalised such that $R=R_{\rm wing}$ is zero for all sight-lines, where $R$ is the distance from the quasar and $R_{\rm wing}$ is the distance at which the damping wing template first exceeds $F=0.01$.  Medians are plotted as the solid lines, and the 68 per cent interval around the median is displayed as the shaded region.  The quasar model corresponds to the blue curves, while the red curves show the same quantities before the quasar has turned on (note again that $x_{\rm HeIII}=0$ initially).  Neutral hydrogen islands are evident at $R-R_{\rm wing}<1\rm\,pMpc$, and elevated gas temperatures due to \HI and \HeII photo-heating by the hard quasar spectrum are present at $R-R_{\rm wing}<0 \rm\,pMpc$ (i.e., behind the \HII and \HeIII ionisation fronts).}
\label{fig:mean_los}
\end{figure*}

\begin{table*}
    \centering
        \caption{Median island properties for islands with $\langle \Delta F \rangle = 0.15 \pm 0.05$.  As in Table~\ref{tab:model_comparison}, the model and quasar properties are listed in columns 1-5.  The fiducial model is in bold.  Column 6 shows the number of lines of sight with $\langle \Delta F \rangle = 0.15 \pm 0.05$, and these are used to construct the median profiles in Figure~\ref{fig:mean_los}.  Columns 7 and 8 give the properties of the median neutral island at the edge of the \HII ionisation front in the stacked lines of sight. These are the size of the neutral island in pMpc, where the neutral island is defined by the length of the continuous region with $x_{\rm HI}> 0.5$, and the logarithm of the \HI column density for the neutral island, calculated over the distance specified in column 7.}
    \begin{tabular}{c|c|c|c|c|c|c|c}
    \hline
        Hydrodynamical  & $\langle x_{\rm HI}\rangle$ &$\rm{M_{1450}}$ & $\rm{\alpha_s}$ & $\rm{t_{\rm q}}$ [yr] &  \# candidates & Island size & $\rm{log_{10}(N_{HI}/\rm cm^{-2})}$   \\
         simulation name & & & &  & & [pMpc] &  \\
        \hline
        {\bf 160-2048-zr53} & {\bf 0.140}  & {\bf -27.4} & {\bf 1.5} & \boldmath{$10^7$} & {\bf 296} & {\bf 0.975} & {\bf 20.01}  \\
        160-2048-zr53 & 0.140 & -27.4 & 0.5 & $10^7$& 332& 1.263& 20.08 \\
        160-2048-zr53& 0.140 &  -27.4 & 2.5 & $10^7$&210 & 0.782& 19.92  \\
        160-2048-zr53 &0.140 & -27.4 & 1.5 & $10^5$  & 485& 1.276 & 20.11\\
        160-2048-zr53 & 0.140 & -27.4 & 1.5 & $10^6$ & 450 & 1.049& 20.03\\
        160-2048-zr53 &0.140  & -27.4 & 1.5 & $10^{7.6}$ & 77 & 0.387 & 19.56\\
        160-2048-zr53  & 0.140  & -27.4 & 1.5 & $10^8$ & 5 & 0.539 & 19.88\\
        160-2048-zr53  & 0.140 & -26.65 & 1.5 & $10^7$ & 230 &0.864 & 19.97\\
        160-2048-zr53 & 0.140 & -28.15 & 1.5 & $10^7$ & 412 & 1.058 & 20.05 \\
        160-2048-zr53 & 0.140 & -28.9 & 1.5 & $10^7$& 493& 1.276 & 20.11 \\
        40-2048-zr53 &0.142 &-27.4 & 1.5 & $10^7$  & 422& 0.243& 19.28\\
        40-2048-zr57 & 0.053 & -27.4 & 1.5 & $10^7$ & 30 & 1.539& 20.12\\
\hline
    \end{tabular}

    \label{tab:neutral island properties}
\end{table*}

We now investigate the underlying physical properties of the simulated spectra that return \deltaF values consistent with J0148.  In Figure \ref{fig:mean_los}, for our fiducial model (left column) and bright quasar model ($M_{1450}=-28.9$, right column), following \citet{Becker24} we take simulated lines of sight with \deltaF $= 0.15 \pm 0.05$ and stack them to find the median and 68 per cent scatter of the underlying physical properties. These are shown in Figure \ref{fig:mean_los} as solid lines and shaded regions respectively.  Note the corresponding \Lya and \Lyb transmission profiles are already displayed in Figure~\ref{fig: wing transmission}. 

From top to bottom, the panels in Figure \ref{fig:mean_los} show the gas density $\Delta = \rho / \langle \rho \rangle$, the \HI fraction, the \HeII fraction, the \HeIII fraction and the gas temperature.   The gas temperature and ionised fractions are furthermore shown just before the quasar emission was turned on (red curves), and for an optically/UV bright lifetime of $t_{\rm q}=10^{7}\rm\,yr$ (blue curves).   Both quasar emission models give similar results. Before the quasar turns on, the lines of sight producing damping wing-like features typically have a large region of neutral hydrogen (initially $\sim \,2$ pMpc) situated in a region that is slightly more underdense than the surrounding IGM.  A very similar result was reported for the simple model used in \cite{Becker24} (see their figure 13), consistent with their suggestion that the J0148 \Lya near zone may terminate at a void.  The neutral region is also significantly cooler  than its partially ionised surroundings because it is yet to be reionised and photo-heated. After the quasar turns on the IGM becomes more highly ionised, and the pre-existing \HII region around the quasar host halo -- originally created by the ionising emission from clustered galaxies -- expands in size.  The quasar ionisation front pushes into the neutral region to ionise the hydrogen and doubly ionise the helium. Note also that prior to the quasar turning on, the regions containing ionised hydrogen contained almost all of their helium in singly ionised form, but for $t_{\rm q}=10^{7}\rm\,yr$ the hard photons from the quasar doubly ionise the helium for $R<R_{\rm wing}$. The temperature in this previously neutral region is increased to $T \sim 4 \times 10^4$K due to the combined effect of \HI and \HeII photoheating \citep{Bolton12}. These increased temperatures also enhance the transmission for $R<R_{\rm wing}$ (since $n_{\rm HI}\propto T^{-0.7}$ in ionisation equilibrium).  This is again consistent with the suggestion in \cite{Becker24} that the \Lya transmission at the edge of the near-zone in the J0148 spectrum may also be associated with gas that has recently been reionised and photoheated. 

Finally, from this stacked data in Figure~\ref{fig:mean_los} we may calculate the typical size of the neutral island at the blue edge of the \Lya near zone.  We define this as the length of the continuous region where $x_{\rm{HI}} \geq 0.5$ ahead of the quasar \HII ionisation front.  Note that stacking allows us to ignore the effects of fluctuations in individual neutral island sizes.  The size of the stacked neutral island and its \HI column density was then calculated, with both properties shown in Table \ref{tab:neutral island properties}.   Typical values in our fiducial model are $\sim 1\rm\,pMpc$ and $N_{\rm HI}\sim 10^{20}\rm\,cm^{-2}$, with the size of the neutral islands across all models similar to or smaller than the maximum extent of the \Lya and \Lyb dark gap in the J0148 spectrum ($7.5h^{-1}\rm\,cMpc$, or $\sim 1.6 \rm\,pMpc$ at $z=6$).


\section{Efficacy of the \deltaF statistic for identifying diffuse neutral islands} \label{sec:efficacy}

\begin{figure}
\centering
    \includegraphics[scale = 0.53]{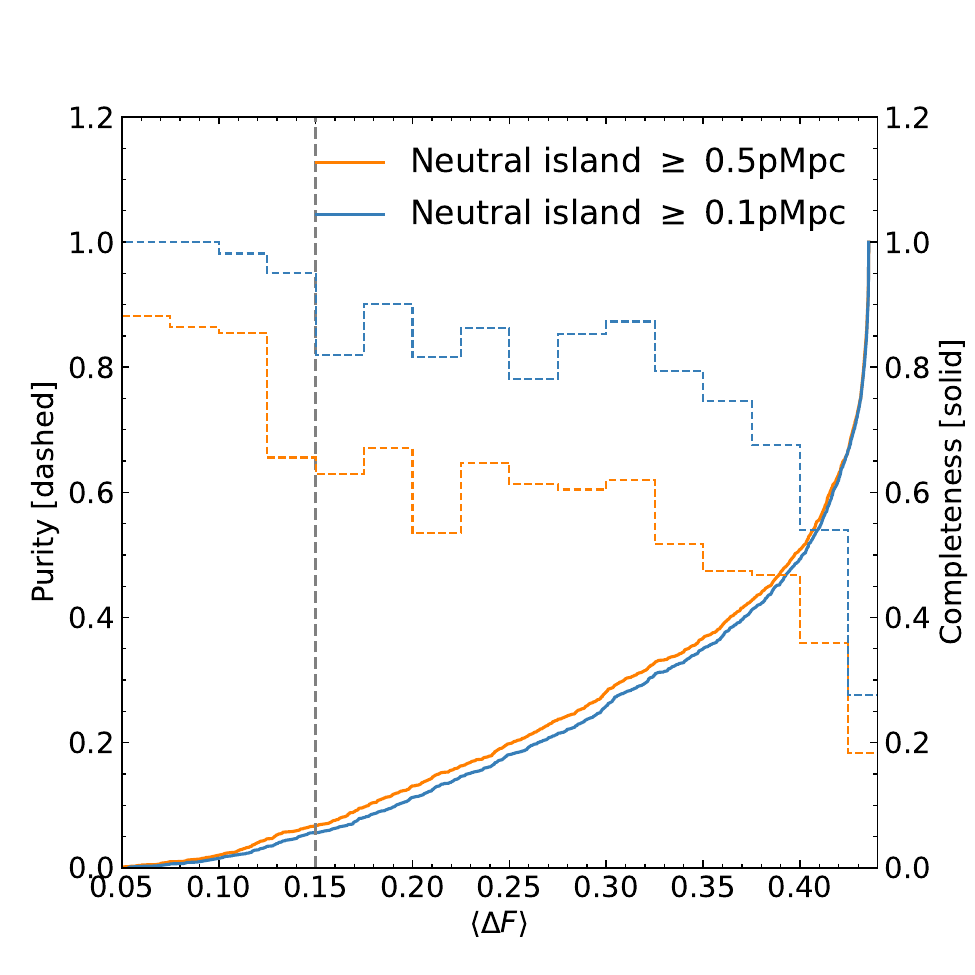}
    \vspace{-0.7cm}
  \caption{Purity and completeness (see text for details) for positive neutral island identification using the $\langle \Delta F \rangle$ statistic introduced by \citet{Becker24}.  The results are shown for our fiducial model, for 6000 individual simulated quasar spectra.  The solid curves show the cumulative distribution of the completeness, while the dashed curves show the differential distribution of the purity.   The orange and blue curves use two different definitions for a neutral island.  The orange curves require that a neutral island must have $x_{\rm HI}>0.5$ over a region of length $\geq 0.5$pMpc to return a positive identification, while the blue curves require a shorter length of $\geq 0.1$pMpc.  The purity for neutral islands with $\geq 0.1$ pMpc and \deltaF $\leq 0.15$ is $0.97$, although the completeness shows these represent only small fraction of the total neutral islands in the simulated quasar spectra.}
    \label{fig:purity-comp}
\end{figure}

We lastly examine the effectiveness of the \deltaF statistic introduced by \citet{Becker24} at correctly identifying diffuse neutral islands in our simulated quasar spectra.  It is expected that the presence of a neutral island close to the edge of the quasar near zone at \rwing will produce a characteristic damping wing shape in the spectrum, and hence return a small value of $\langle \Delta F \rangle$.  However, a similar signal could be produced by chance without the presence of a neutral island.   

To assess the presence of neutral islands in the diffuse IGM near \rwing we must first adopt an operational definition for a neutral island.  We opt to select regions where $x_{\rm{HI}} \geq 0.5$ continuously over a distance of either $0.1$ or $0.5\rm\, pMpc$, where addditionally the red edge of the neutral island must fall between \rwing and $R_{\rm wing} + 1\rm\,pMpc$.   Multiple choices for the position of the red edge of the neutral island were tested, but this choice was not found to affect our results significantly; most neutral islands that produce a damping wing have a red edge very close to $R_{\rm wing}$.  However, the distance over which we require a neutral region to have $x_{\rm HI}\geq 0.5$ has a larger effect. In our analysis we therefore use both $0.1$ or $0.5\rm\, pMpc$.   Adopting the larger value of $0.5\rm\,pMpc$ excludes sight-lines that exhibit multiple small neutral islands that are punctuated by narrow regions of highly ionised IGM. 

Next, using our 6000 simulated lines of sight drawn from the fiducial model, we may calculate both the purity and completeness of the neutral island sample recovered from the simulated spectra as a function of the \deltaF statistic.  We define the purity as
\begin{equation} {\rm Purity} =  \frac{N({\rm neutral\, island} \land x \leq \langle \Delta F \rangle < x+\delta x)}{N(x \leq  \langle \Delta F \rangle < x+\delta x)}, \end{equation}
where $N({\rm neutral\, island} \land x \leq \langle \Delta F \rangle < x+\delta x)$ is the number of lines of sight with \deltaF between $x$ and $x+\delta x$ that have a neutral island, following the definition above. We adopt a constant bin size of $\delta x= 0.025$ in our analysis. 
The completeness is instead defined as a cumulative distribution, where
\begin{equation} {\rm Completeness} = \frac{N({\rm neutral \, island} \land \langle \Delta F \rangle \leq x)}{N({\rm neutral\, island})}. \end{equation} 
Here $N({\rm neutral\, island} \land \langle \Delta F \rangle \leq x)$ is the number of lines of sight that contain neutral islands and have \deltaF $\leq x$. 

The purity and completeness of the simulated quasar spectra in our fiducial model are presented in Figure \ref{fig:purity-comp} for our two choices of neutral island length.  Note again that we do not model damping wings from compact absorbers in our simulations, and so any effect these may have on the results is not quantified here.   For neutral islands with size $\geq 0.1$ pMpc, the purity of all lines of sight with \deltaF $\leq 0.15$ is $0.97$, although this value is lowered to $0.78$ if restricting our definition to include only continuous neutral islands with size $\geq 0.5$ pMpc.  This implies a non-negligible fraction of the simulated spectra have damping wings that arise from multiple small neutral islands separated by narrow ionised regions.  In general, therefore, the purity will be larger for smaller $\langle \Delta F \rangle$, but will only reach unity for $\langle \Delta F \rangle \leq 0.15$.   The completeness (solid curves) is already non-zero at $\langle \Delta F \rangle<0.1$ but quickly reaches unity at $\langle \Delta F \rangle > 0.4$, demonstrating there are many lines of sight containing neutral islands that do not exhibit a \Lya damping wing profile as obvious as J0148 (an example is shown in the right column of Figure~\ref{fig:example_los}).

In summary -- assuming that damping wings from compact absorbers can be ruled out -- we confirm the method introduced by \citet{Becker24} should reliably indicate the presence of a neutral island of extent $>0.1\rm\,pMpc$ in the spectrum of J0148 if \deltaF $\leq 0.15$ (i.e. if there is a reasonably obvious damping wing profile).  However, we also expect this criterion will only catch at most a few per cent of the total population of neutral islands in the vicinity of quasar near-zones.  The detection of damping wings from neutral islands in stacked spectra \citep[e.g.][]{MalloyLidz2015,Zhu24,Spina24} may be a more effective approach in these cases.  Alternatively, the detection of $21\rm\,cm$ absorption from the diffuse IGM at $z\simeq 6$ could also provide evidence for the presence of neutral islands, although the detectability will depend on the strength of the heating in the neutral IGM by the X-ray background \citep{Soltinsky2021,Soltinsky2024}.


\section{Conclusions} \label{sec: conclusions}

\cite{Becker24} recently demonstrated the shape of the \Lya transmission in the near zone of the  $z=5.9896$ quasar ULAS J0148$+$0600 (hereafter J0148) is consistent with the damping wing profile expected from an extended neutral hydrogen island in the diffuse intergalactic medium (IGM).  This represents some of the first direct evidence for an extended region of neutral hydrogen in the diffuse, low density IGM at $z<6$ \citep[see also][]{Zhu24,Spina24}.  

In this work, we therefore use simulations of late ending reionisation from Sherwood-Relics \citep{Puchwein23}, combined with 1D radiative transfer simulations of the ionising radiation from the quasar \citep{Bolton07,Soltinsky23}, to better quantify the expected incidence rate of quasar sight-lines with \Lya and \Lyb absorption that quantitatively match the J0148 spectrum.  We also use our simulations to investigate the underlying physical properties of the neutral island implied by J0148 damping wing, and test the efficacy of the summary statistic, $\langle \Delta F \rangle$ introduced by \citet{Becker24} for identifying diffuse neutral hydrogen in the IGM.   

We confirm that simulations with reionisation ending late at $z=5.3$ \citep[see also][]{Kulkarni19,Keating20a,Keating20b,Bosman22} are necessary for reproducing quasar spectra at $z\leq 6$ that exhibit a damping wing profile at the blue edge of the quasar \Lya near zone consistent with J0148.   Although quasar spectra with \deltaF matching J0148 are rare ($p(\langle \Delta F \rangle\leq 0.15)\sim 0.03$) in our late reionisation model,  almost all ($97$ per cent) of our simulated quasar spectra with $\langle \Delta F \rangle\leq 0.15$ contain a neutral island of extent $>0.1\rm\,pMpc$ blueward of the \Lya near zone.  Hence, assuming the damping wing in the spectrum of J0148 does not originate from a compact absorber \citep[see][for a detailed discussion of this point]{Becker24}, it appears highly likely the damping wing reported by \citet{Becker24} with $\langle \Delta F \rangle=0.15$ arises from a neutral hydrogen island in the underdense IGM with approximate extent of $\sim 1\,\rm pMpc$ and total \HI column density $N_{\rm HI}\sim 10^{20}\rm\,cm^{-2}$.  Our modelling furthermore predicts the \Lya transmission at the edge of the \Lya near zone should consist of recently reionised hydrogen and helium with temperature $T\sim 4\times 10^{4}\rm\,K$, as suggested by \citet{Becker24}. 
 
Intriguingly, however, we also find that the a priori probability of drawing a quasar line of sight with both a damping wing profile \emph{and} \Lya near zone size (defined here as $R_{\rm wing}=7\rm\,pMpc$) that matches J0148 is low, $p<10^{-2}$, even when allowing for an optically/UV bright quasar lifetime of up to $t_{\rm q}=10^{8}\rm\,yr$.   The main difficulty is reproducing a region of sufficiently high transmission at the edge of the \Lya near zone, given the photo-ionisation rate from the quasar will decline at least as rapidly as $\Gamma_{\rm HI}\propto R^{-2}$.  Matching the J0148 spectrum requires highly underdense gas and significantly elevated gas temperatures at the near-zone edge, and this combination occurs relatively infrequently in our models.   The probability of recovering a J0148-like spectrum from our models can be modestly increased if we assume J0148 produces significantly more ionising photons than expected for standard assumptions (e.g., for a harder extreme UV spectrum, $\alpha_{\rm s}<1.5$ or a brighter UV absolute magnitude, $M_{1450}<-27.4$).  However, this would require the composite quasar continuum for J0148 and/or the quasar UV photometry to be significantly in error.

It remains possible that J0148 is simply an unusual line of sight; it is the only object to date at $z<6$ which exhibits evidence for both a damping wing and a giant \Lya absorption trough. We nevertheless speculate that there could be some additional physical effects missing from our models.  One possibility is quasar variability.  If J0148 has very recently dimmed on a timescale of $t<t_{\rm eq}\sim 10^{5}\rm\,yr$, where $t_{\rm eq}$ is the \HI equilibriation timescale in the quasar near zone, this might explain the apparent tension between the observed near zone extent and the quasar UV luminosity.  Alternatively, as suggested by \citet{Becker24}, the \Lya transmission at the blue edge of the J0148 near zone may represent evidence for the transverse proximity effect; ionising sources close to the quasar line of sight may provide the additional ionising photons needed.   Indeed, \citet{Eilers24} have recently reported a group of \OIII-emitting galaxies just redward of the J0148 near zone. Undetected quasars may also play a role.   These should cluster on the relevant scales but may not be detectable due to a narrow opening angle or variability.  Their hard spectra could furthermore  assist with enhancing gas temperatures at the near-zone edge by doubly ionising helium.    We suggest that further investigation of these issues may be of interest.  Finally, observational estimates of the IGM gas temperature from the thermal broadening of the \Lya transmission at the edge of the J0148 near zone will be valuable.  If a neutral island has recently been reionised by J0148, we predict the IGM temperature should be $T\sim 4\times 10^{4}\rm\,K$ within a few proper Mpc of the \Lya near zone edge.  Direct evidence for enhanced gas temperatures due to the recent photo-ionisation of a neutral island would provide further confirmation that we are witnessing the final stages of reionisation at $z<6$.

\section*{Acknowledgements}

The simulations used in this work were performed using the Joliot Curie supercomputer at the Tr{\'e}s Grand Centre de Calcul (TGCC) and the Cambridge Service for Data Driven Discovery (CSD3), part of which is operated by the University of Cambridge Research Computing on behalf of the STFC DiRAC HPC Facility (www.dirac.ac.uk).  We acknowledge the Partnership for Advanced Computing in Europe (PRACE) for awarding us time on Joliot Curie in the 16th call. The DiRAC component of CSD3 was funded by BEIS capital funding via STFC capital grants ST/P002307/1 and ST/R002452/1 and STFC operations grant ST/R00689X/1.  This work also used the DiRAC@Durham facility managed by the Institute for Computational Cosmology on behalf of the STFC DiRAC HPC Facility. The equipment was funded by BEIS capital funding via STFC capital grants ST/P002293/1 and ST/R002371/1, Durham University and STFC operations grant ST/R000832/1. DiRAC is part of the National e-Infrastructure.  JSB and LC are supported by STFC consolidated grant ST/X000982/1.  Support by ERC Advanced Grant 320596 ‘The Emergence of Structure During the Epoch of Reionization’ is gratefully acknowledged. MGH has been supported by STFC consolidated grants ST/N000927/1 and ST/S000623/1.  We thank Volker Springel for making P-Gadget-3 available. We also thank Dominique Aubert for sharing the ATON code, and Philip Parry for technical support.

\section*{Data Availability}
All data and analysis code used in this work are available from the first author on reasonable request.  Further guidance on accessing the publicly available Sherwood-Relics simulation data may also be found at \url{https://www.nottingham.ac.uk/astronomy/sherwood-relics/}.



\bibliographystyle{mnras}
\bibliography{bibliography.bib} 

\newpage
\appendix

\section{The size of the velocity window used to calculate \deltaF}
\label{app:region size}

\begin{figure*}
    \includegraphics[width = \linewidth]{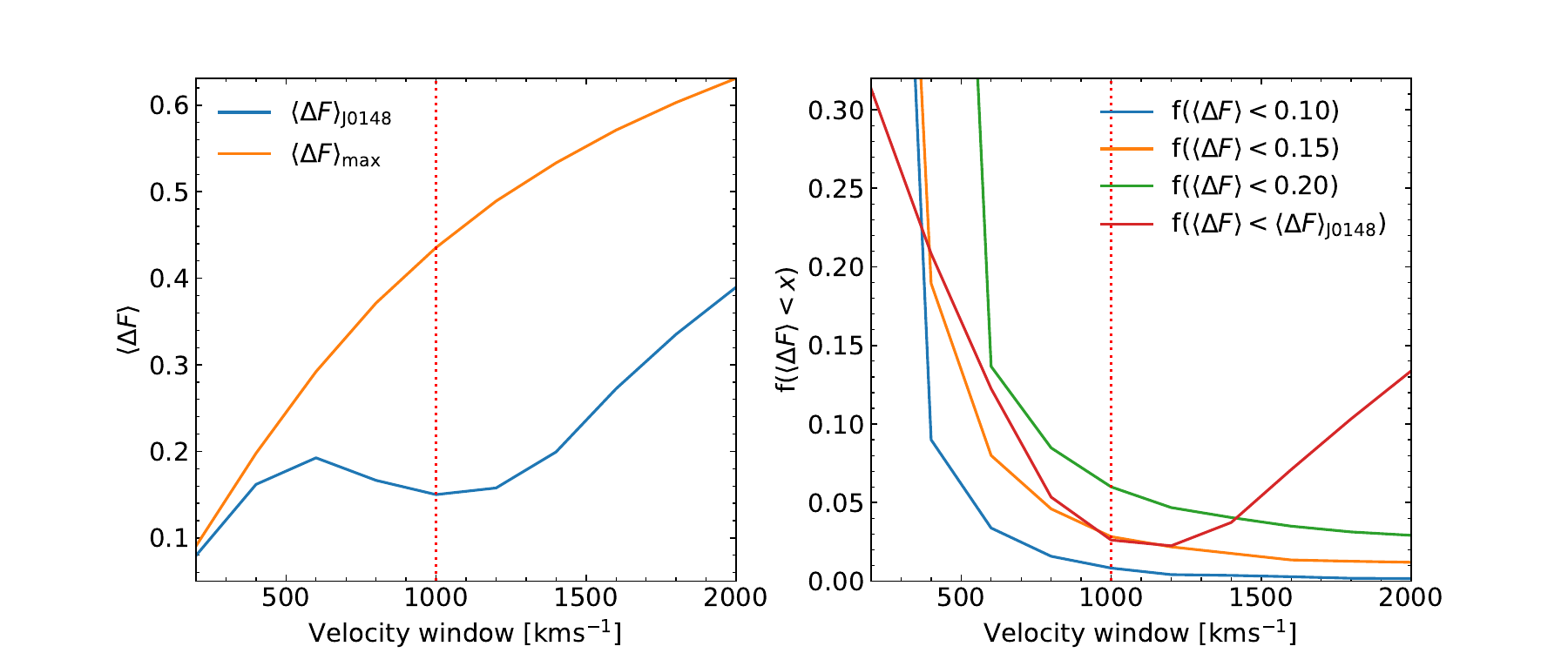}
    \vspace{-0.5cm}
    \caption{Left: The \deltaF statistic against the size of the velocity window used to calculate it. In blue is the \deltaF value calculated for J0148 as a function of the window size in $\rm{km\,s^{-1}}$, and in orange is the maximum value \deltaF may take over this window (due to the damping wing shape, this is lower for smaller velocity windows). The dotted red line indicates the 1000 $\rm{kms^{-1}}$ velocity window used throughout our work. Right: the fraction of sightlines, $f(\langle \Delta F \rangle < x)$, in our fiducial model with \deltaF below some value $x$. The blue, orange, and green curves show three fixed values of $x=0.10$, $0.15$ and $0.20$, respectively, while the red line shows the probability that \deltaF is lower than the observed J0148 value (shown by the blue curve in the left panel).}
    \label{fig:regionsize}
    \end{figure*}
We have chosen a $1000\, \rm{km\,s^{-1}}$ velocity window to calculate \deltaF, following \citet{Becker24}. Here we briefly assess the sensitivity of our results to this choice.

The blue curve in the left panel of Figure~\ref{fig:regionsize} shows the \deltaF value obtained from the spectrum of J0148 as a function of the velocity window size.  For comparison, the orange curve shows the maximum possible value \deltaF can take.  The first point to note is that the velocity window must be large enough to reliably assess the presence of a damping wing.  If the velocity window is made too small, $\lesssim 500\rm\,km\,s^{-1}$, the measured \deltaF is almost always close to the maximum. This is due to the saturated absorption that is present at the edge of the near zone (both in the simulated and observed spectra).  The summary statistic then ceases to be informative.  Another consideration when choosing the size of the velocity window is the avoidance of the region of saturated absorption in the spectrum of J0148 (see the grey spectrum in Figure \ref{fig: wing transmission}, starting at $R - $\rwing$ = -2$  pMpc). It can be seen in the left panel of Figure \ref{fig:regionsize} that for windows $> 1200\, \rm{km\,s^{-1}}$, $\langle \Delta F \rangle_{\rm J0148}$ increases as a consequence of this saturated region.  One might therefore ask whether adopting a larger velocity window that includes this absorption, say $\sim 1500\rm\, km\,s^{-1}$, will change our conclusions regarding the rarity of the J0148 configuration.

This is addressed in the right panel of Figure \ref{fig:regionsize}, which shows the fraction of simulated sight-lines, $f(\langle \Delta F \rangle < x)$, in our fiducial model with $\langle \Delta F\rangle < x$ as a function of the velocity window size.   There is a change of a factor two in $f(\langle \Delta F \rangle < \langle \Delta F \rangle_{\rm J0148})$ from 1000 $\rm{km\,s^{-1}}$ to 1500 $\rm{km\,s^{-1}}$, where $\langle \Delta F \rangle_{\rm J0148}$ corresponds to the blue curve in the left panel of Figure \ref{fig:regionsize}.  Quantitatively, the probability of the J0148 configuration changes from $p=0.03$ for 1000 $\rm{kms^{-1}}$ in Table \ref{tab:model_comparison} to $p=0.06$ for 1500 $\rm{kms^{-1}}$.  However, as discussed in the main text, it is the simultaneous reproduction of \deltaF and \rwing that makes this particular sight-line rare. By this metric (column 7 in Table \ref{tab:model_comparison}), we obtain $p=0.0002$ for 1000 $\rm{km\,s^{-1}}$ and $p=0.0013$ for 1500 $\rm{km\,s^{-1}}$ with our fiducial model, still a very rare occurrence. In summary, our choice for the size of the velocity window used to compute \deltaF does not change our conclusion that J0148 is a rare configuration in our models.

\section{Algorithm for fitting the damping wing template} \label{app:flowchart}
\begin{figure*}
    \centering
    \includegraphics[trim={12cm 2cm 0cm 0cm},clip, scale = 0.65]{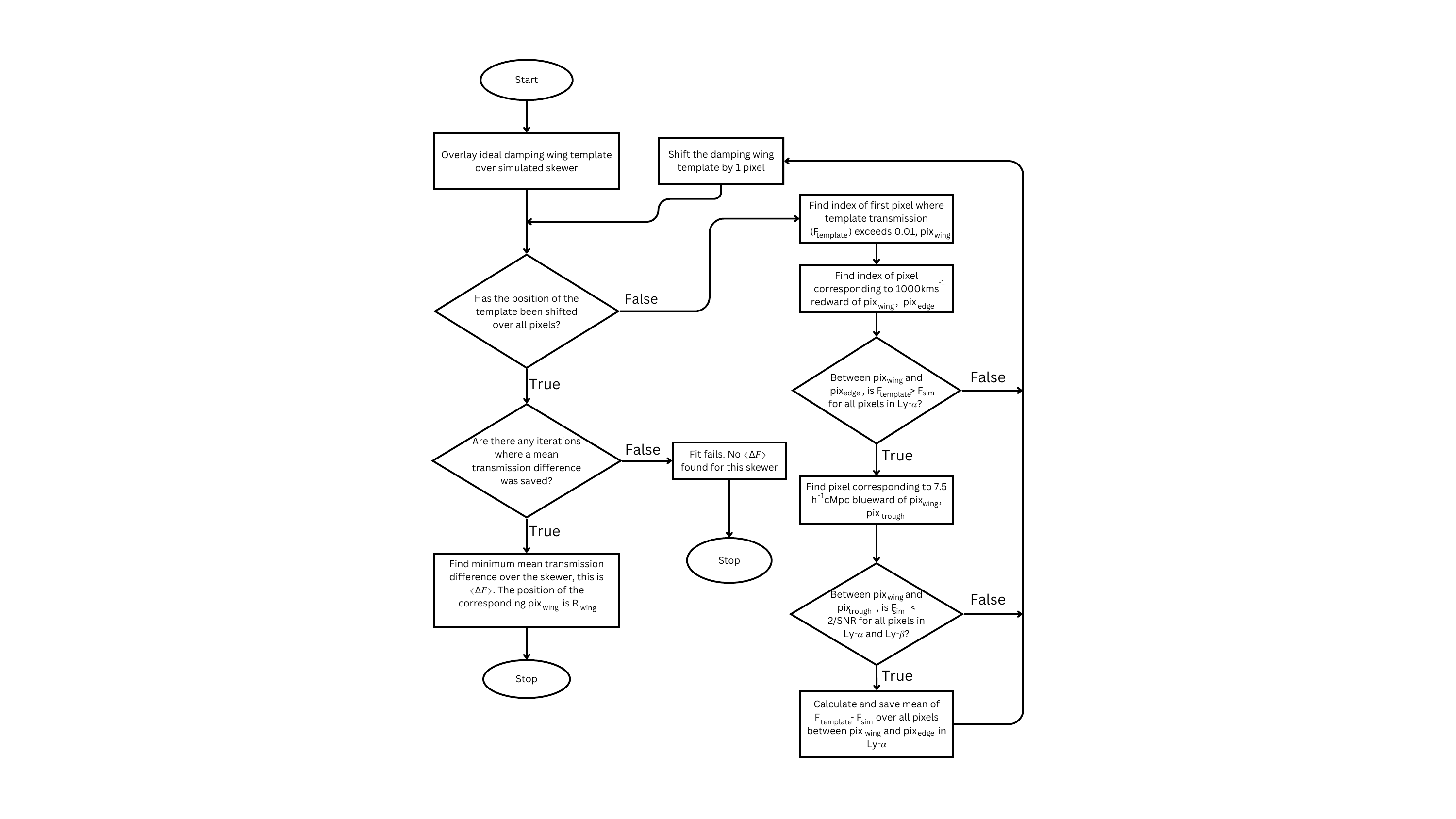}
    \vspace{-0.4cm}
    \caption{A flow chart outlining the algorithm used in this work to find  \deltaF and $R_{\rm wing}$ for each simulated quasar line of sight (see also Section~\ref{sec:idealwing} and \citet{Becker24} for further details).}
    \label{fig:flowchart}
\end{figure*}

In Section~\ref{sec:idealwing} we described the procedure we use for fitting the damping wing template to our simulated spectra \citep[see also][]{Becker24}.   The flow chart displayed in Figure~\ref{fig:flowchart} provides a graphical summary of this algorithm.

\section{Numerical convergence with mass resolution} 
\label{app:converge}

\begin{figure}
    \centering
    \includegraphics[trim={0cm 0cm 0cm 0cm},clip, width=\linewidth]{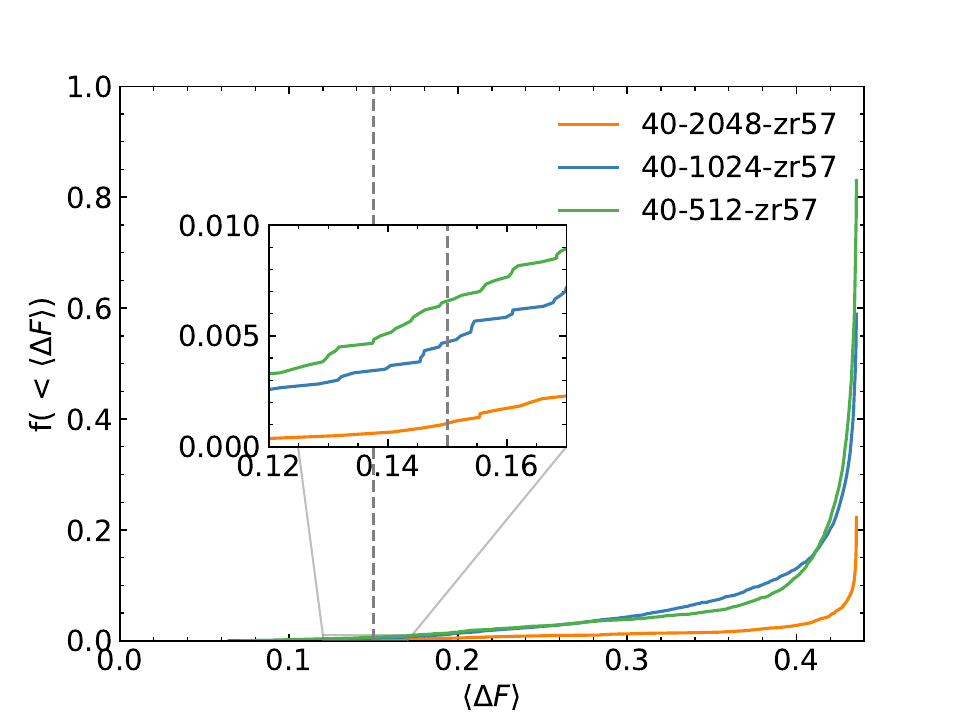}
    \vspace{-0.7cm}
    \caption{As for Figure~\ref{fig:CDFs}, but now for simulations in a fixed box size of $40h^{-1}\rm\,cMpc$ with different mass resolutions: 40-2048-zr57 (orange curve), 40-1024-zr57 (blue curve) and 40-512-zr57 (green curve). The grey dashed line shows \deltaF = 0.15.  The inset shows a zoom-in of the region around \deltaF = 0.15.}
    \label{fig:convergence-test}
\end{figure}

We use two additional simulations with reduced mass resolution in $40h^{-1}\rm\,cMpc$ boxes to test numerical convergence; 40-512-zr57 with gas particle mass $M_{\rm gas}=6.38\times 10^{6}h^{-1}\rm\,M_{\odot}$ and 40-1024-zr57 with $M_{\rm gas}=7.97\times 10^{5}h^{-1}\rm\,M_{\odot}$.   The simulations are otherwise identical to 40-2048-zr57 (see Table~\ref{table:sim_params}).   In Figure \ref{fig:convergence-test}, we show $f(<\langle \Delta F \rangle)$ for each model.  Note also the 40-512-zr57 simulation and our fiducial 160-2048-zr53 run have the same mass resolution. 

The inset in Figure \ref{fig:convergence-test} shows the zoomed-in region around \deltaF = 0.15. The absolute difference between the different models at \deltaF $= 0.15$ is small, with $f(<\langle \Delta F \rangle)<0.007$ for all models.   These differences remain small compared to the change from other factors such as the quasar absolute magnitude or optically/UV bright lifetime.  The differences at $\langle \Delta F \rangle>0.2$ are much larger.  This is because the lower resolution models produce smoother transmission, and so the requirement that the simulated spectrum does not exceed the ideal wing template over our calculation region is more likely to be satisfied.  This also imples that if the mass resolution were increased further, J0148-like spectra would be even rarer in our models.



\bsp	
\label{lastpage}
\end{document}